\documentclass[%
 reprint,
 amsmath,amssymb,
 aps,prc
]{revtex4-2}
\usepackage[colorlinks, citecolor=red]{hyperref}
\usepackage{graphicx}
\usepackage{epstopdf}
\usepackage{dcolumn}
\usepackage{bm}
\usepackage{natbib}
\usepackage{multirow}
\begin{document}
\preprint{APS/123-QED}
\title{Kaon Meson Condensation of Hyperonized Neutron Star within the framework of the $\sigma$\mbox{-}cut Scheme}

\author{Fu Ma$^{1}$, Wenjun Guo$^{1}$ and Chen Wu$^{2}$\footnote{Electronic address: wuchenoffd@gmail.com}} \affiliation{
\small 1. University of Shanghai for Science and Technology, Shanghai 200093, China\\
\small 2. Shanghai Advanced Research Institute, Chinese Academy of Sciences, Shanghai 201210, China}

\begin{abstract}
The recent measurement of the mass of neutron stars (PSR J1614 - 2230, PSR J0348 + 0432, MSP J0740 + 6620) restricts the lower limit $\sim 2M_{\odot}$ of the maximum mass of such compact stars, making it possible for dense matter  to exist in massive stars.  The relativistic mean field theory with parameter sets FSUGold including Kaon condensation is used to describe the properties of neutron stars in $\beta$ equilibrium. Through careful choice of the parameter of the $\sigma$-cut $c_{\sigma}$, we are able to produce a maximum mass neutron star with Kaon condensation heavier than $2M_{\odot}$, and we find that the parameter $\Lambda_{\nu}$ of the $\rho-\omega$ interaction term in this model has a significant effect on $K^{-}$ condensation. In the case of using $\sigma$-cut scheme, $K^{-}$ condensation occurs only when the $\rho-\omega$ interaction $\Lambda_{\nu}$ is switched off.
\end{abstract}

\maketitle

\section{\label{sec:level1}INTRODUCTION},

Born as a result of supernova explosions, neutron stars are highly condensed stellar remnants. A typical neutron star has a mass of the order of the solar mass, but its radius is only around 10$\sim$12 km. The internal matter composition and equation of state (EOS) of compact stars are restricted by recent gravitational wave observations. The massive NSs observed,e.g.,PSR J1614-2230 with $M=1.908\pm0.016M_{\odot}$ \cite{Demorest:2010bx,arzoumanian2018nanograv,fonseca2016nanograv,ozel2010massive},have established strong constraints on the EOS of nuclear matter. PSR J0348+0432 with $M=2.01\pm0.04M_{\odot}$ \cite{antoniadis2013j}, MSP J0740+6620, with $M=2.08^{+0.07}_{-0.07}M_{\odot}$ \cite{fonseca2021refined,cromartie2020relativistic}, and radius $12.39^{+1.30}_{-0.98}$ km obtained from NICER data \cite{riley2021nicer}, have strengthened the already stiff constraints on the EOS. The observation of gravitational waves is another crucial source of information about NS matter, the recent GW190814 event observed by the LIGO - Virgo Collaboration (LVC) from a coalescence of a black hole and a lighter companion sets the mass of the former to be $23.2_{-1.0}^{+1.1}$ $M_{\odot}$ and that of the latter to be $2.59_{-0.09}^{+0.08}$ $M_{\odot}$ \cite{abbott2020gw190814}. It is unclear whether the lighter companion star is a neutron star or a black hole. So far, the study of neutron stars has been used as one of the important methods for studying strong interactions in dense nuclear matter, and it is still a research hotspot of nuclear celestial bodies \cite{glendenning2012compact,Weber:2006ep}. It has gradually been discovered that only considering basic nucleon, neutrons and protons is not enough in the study of neutron stars.
 In fact, the inner cores of the neutron stars are sources of speculation, some of the possibilities being the appearance of hyperons \cite{Schaffner:1995th,Wu:2011zzb}, only quarks 
\cite{Pal:1999sq,Glendenning:1992kd,Glendenning:1993vx,Li:2011vd,Li:2010gx}, or Kaon condensates \cite{Thapa:2020usm,Maruyama:2005tb,Brown:2005yx,Shao:2010zz}.  
\par{}
Kaplan and Nelson have suggested that the ground state of hadronic matter might form a negatively charged Kaon Bose-Einstein condensation above a certain critical density \cite{Kaplan:1986yq,Nelson:1987dg}. In the interior of a neutron star, as the density of neutrons increases, the electronic chemical potential will increase to keep the matter in $\beta$-equilibrium. When the electronic chemical potential exceeds the mass of muons, muons appear. And when the vacuum mass of the meson (pion, Kaon) is exceeded, as the density increases, negatively charged mesons begin to appear, which helps to maintain electrical neutrality. However, the $s$ - wave $\pi N$ scattering potential repels the ground state mass of the $\pi$ meson and prevents the generation of the $\pi$ meson \cite{Glendenning:1984jr}. The effective mass of the $K^{-}$ meson is decreased due to the interaction with the nucleon. If the $K^{-}$ meson energy intersects with the electron chemical potential at a certain density, then $K^{-}$ will be more advantageous than electrons as a neutralizer for positive charges.The very interaction that reduces the Kaon energy modifies the nucleons with which they interact, and this will open the possibility of the appearance of Kaon condensates.
\par
Although many scholars have proposed various phenomenological models based on density functional theory and realistic nuclear potential to explain, the interaction of particles at such high densities is not known precisely. To get the EOS of the neutron star matter, the RMF theory is usually applied \cite{Walecka:1974qa}, which describes the interaction between baryons via the exchange of mesons \cite{Nelson:1987dg,Vautherin:1971aw,Shen:1998gq}. The model determines the coupling parameters through the saturation properties of the nuclear matter and extends it to high-density.  It has achieved great success in the study of the nuclei and nuclear matter. There are many theoretical models within the RMF framework. In this paper, we are going to adopt the FSUGold model as an example \cite{ToddRutel:2005zz}. The RMF theory with the parameter set FSUGold was proposed by Todd-Rutel and Piekarewicz \cite{Fattoyev:2010mx,Piekarewicz:2007us,Piekarewicz:2007dx,Fattoyev:2010rx}. It has achieved great success in study of the nuclear matter and the ground-state properties of some spherical nuclei \cite{Piekarewicz:2007dx}. However, when it is applied to study the EOS including hyperons, the maximum mass of the neutron stars obtained by this model is too small, the problem is that the EOS generated by the FSUGold model \cite{Wu:2011zzb} is too soft. In order to describe nuclear matter at high density with FSUGold model, some researchers have proposed a $\sigma$-cut scheme \cite{Maslov:2015lma}. This scheme points out that if the mean-field self-interaction potential rises sharply in a narrow vicinity $n_{0}$$\sim$$n$ of the value of mean fields corresponding to nuclear densities $n_{*}$ \cite{Maslov:2015lma}, the nucleon effective mass saturates and the EOS stiffens, where $n_{0}$ is the nuclear saturation density. This procedure offers a simple way to stiffen the EOS at high densities without altering it at densities $n \leq n_{0}$. However, the effect of hyperons on neutron star matter is not included in study \cite{Maslov:2015lma}. It is well known that the introduction of hyperon degrees of freedom leads to a softer EOS, thus producing a neutron star with a small mass. It leaves open question of whether the $\sigma\mbox{-}$ cut scheme is still effective when hyperons are considered. In addition, some scholars pointed out that exotic EOS can not be ruled out by the observation of a $2M_{\odot}$ compact star \cite{Char:2014cja}. In this article, we use the FSUGold model to study NS matter including hyperons and Kaon condensates with $\sigma \mbox{-}$cut scheme.
\par
This paper is organized as follows. First, the theoretical framework is presented. Then we will study the effects of the $\sigma$-cut scheme. The Kaon condensation will be considered too. Finally, some conclusions are provided. 

\section{\label{sec:level2}theoretical framework}

In this section, we introduce the FSUGold model to study the properties of the phase transition from hadronic to Kaon condensed matter. For the baryons matter we have considered nucleons ($n$ and $p$), and hyperons ($\Lambda, \Sigma$ and $\Xi$), The exchanged mesons include the isoscalar scalar meson ($\sigma$), the isoscalar vector meson ($\omega$), the isovector vector meson ($\rho$), the starting point of the extended FSUGold model is the Lagrangian density:
\begin{widetext}
\begin{equation}
\begin{aligned}
\mathcal{L}=&\sum_{B}\bar{\psi}_{B}[\emph{$i$}\gamma^{\mu}\partial{\mu}-m_B+g_{\sigma B}\sigma-g_{\omega B}\gamma^{\mu}\omega_{\mu}-\frac{g_{\rho B}}{2}\gamma^{\mu}\vec{\tau}\cdot\vec{\rho^{\mu}}]\psi_{B}+\frac{1}{2}\partial_{\mu}\sigma\partial^{\mu}\sigma\\
&-\frac{1}{2}m_{\sigma}^{2}\sigma^{2}-\frac{\kappa}{3!}(g_{\sigma N}\sigma)^3-\frac{\lambda}{4!}(g_{\sigma N}\sigma)^{4}-\frac{1}{4}F_{\mu \nu}F^{\mu \nu}+\frac{1}{2}m_{\omega}^{2}{\omega_{\mu}}\omega^{\mu}+\frac{\xi}{4!}(g_{\omega N}^{2}\omega_{\mu}\omega^{\mu})^{2}\\
&+\frac{1}{2}m_{\rho}^{2}\vec{\rho}_{\mu}\cdot\vec{\rho}^{\mu}-\frac{1}{4}\vec{G}_{\mu \nu}\vec{G}^{\mu \nu}+\Lambda_{\nu}(g_{\rho N}^{2}\vec{\rho}_{\mu}\cdot\vec{\rho}^{\mu})(g_{\omega N}^{2}\omega_{\mu}\omega^{\mu})+\sum_{l} \bar{\psi}_{l}[i{\gamma}^{\mu}\partial{\mu}-m_{l}]{\psi}_{l},
\label{eq:one}
\end{aligned}
\end{equation}
\end{widetext}

where $\Lambda_{\nu}$ is introduced to modify the density dependence of symmetry energy. All the eight lightest baryons ($\emph{p,n},\Lambda^{0},\Sigma^{+},\Sigma^{0},\Sigma^{-},\Xi^{0},\Xi^{-}$) are included, as the two leptons, electron and muon. The terms $\emph{m}_{\sigma}$, $\emph{m}_{\omega}$, and $\emph{m}_{\rho}$ are the masses of $\sigma$, $\omega$, and $\rho$ mesons, respectively. The antisymmetric tensors of vector mesons take the forms $\emph{F}_{\mu \nu}=\partial_{\mu}\omega_{\nu}-\partial_{\nu}\omega_{\mu}$, $\emph{G}_{\mu \nu}=\partial_{\mu}\rho_{\nu}-\partial_{\nu}\rho_{\mu}$. The isoscalar meson self-interactions (via $\kappa$, $\lambda$ and $\xi$ terms) are necessary for the appropriate EOS of symmetric nuclear matter. $g_{\sigma N}$, $g_{\omega N}$ and $g_{\rho N}$ are the coupling constants between baryon and $\sigma$ meson, baryon and $\omega$ meson and baryon and $\rho$ meson, respectively. In this paper, the operators of meson fields are replaced by their expectation values by the mean field approximation.

With the increase of the density, the Kaon condensation does appear in the interior of the neutron stars. We take the Lagrangian of Kaon condensation as the same that is Ref. \cite{Glendenning:1997ak} and \cite{Glendenning:1998zx}, which reads
\begin{equation}
{\cal{L}}_{K}=D_{\mu}^{*}K^{*}D^{\mu}K-m_{K}^{* 2}K^{*}K,
\label{eq:two}
\end{equation}
where $\emph{D}_{\mu}=\partial_{\mu}+ig_{\omega K}\omega_{\mu}+i\frac{g_{\rho K}}{2}\tau_{\emph{K}}\cdot\rho_{\mu}$ is the covariant derivative and the Kaon effective mass is defined as 
$\emph{m}_{\emph{K}}^{*}=\emph{m}_{\emph{K}}-g_{\sigma K}\sigma$. The coupling constants between the vector meson and the Kaon $\emph{g}_{\omega K},g_{\rho K}$ are determined by the meson SU(3) symmetry as $g_{\omega K}=g_{\omega N}/3,g_{\rho K}=g_{\rho N}$ \cite{Char:2014cja}. The scalar coupling constant $g_{\sigma K}$ is fixed to the optical potential of the $\emph{K}^{-}$ at saturated nuclear matter:
\begin{equation}
\label{eq3}
\emph{U}_{K}(\rho_{0})=-g_{\sigma K}\sigma(\rho_{0})-g_{\omega K}
\omega(\rho_{0}),
\end{equation}
which characterizes the Kaon-nucleon interaction. Experiment studies show that the kaons experience a repulsive interaction in nuclear matter whereas antikaons experience an attractive potential \cite{Li:1997zb,Pal:2000yc}. Waas and Weise found an attractive potential for the $\emph{K}^{-}$ at the saturation nuclear density of about $\emph{U}_{K}(\rho_{0})=-120 MeV$ \cite{Waas:1997pe}. Coupled channel calculations at finite density have yielded a value of $\emph{U}_{K}(\rho_{0})=-100$ MeV \cite{Koch:1994mj}. More recent self-consistent calculations with a chiral Lagrangian \cite{Lutz:1997wt,Ramos:1999ku} and coupled channel calculation including a modified self-energy of the Kaon \cite{Tolos:2002ud} indicate that the Kaon may experience an attractive potential with its depth about -80 MeV to even -50 MeV at the saturation density. Another calculation from hybrid model \cite{Friedman:1998xa} suggests the value of $K^{-}$ optical potential to be in the range 180 $\pm$ 20 MeV at saturation density. In this paper, we carry out our calculations with a series of optical potentials ranging from - 160 MeV to -120 MeV.
\par
For the meson-hyperon couplings, we take those in the SU(6) quark model for the vector couplings constants:
\begin{equation}
g_{\rho \Lambda}=0, g_{\rho \Sigma}=2g_{\rho \Xi}=2g_{\rho N},
\end{equation}
\begin{equation}
g_{\omega \Lambda}=g_{\omega \Sigma}=2g_{\omega \Xi}=\frac{2}{3}g_{}{\omega N}.
\end{equation}
The scalar couplings are usually fixed by fitting hyperon potentials with $U_{Y}^{(N)}=g_{\omega Y}\omega_{0}-g_{\sigma Y}\sigma_{0}$, where $\sigma_{0}$ and $\omega_{0}$ are the values of the scalar and vector meson strengths at saturation density \cite{Schaffner:1993qj}. The $\Lambda\mbox{-}N$ interaction has been well studied and $U_{\Lambda}^{N}=-30$ MeV was obtained with bound $\Lambda$ hypernuclear states \cite{Millener:1988hp,Malik:2021nas}. One of the unsettled issues in hypernuclear physics is the $\Sigma\mbox{-}N$ interaction in nuclear matter. An attractive potential was generally used in the past for $\Sigma$ to be bounded in nuclear matter. However, a detailed scan for $\Sigma$ hypernuclear states turned out to give negative results. The study of $\Sigma^{-}$ atoms also showed strong evidence for a sizable repulsive potential in the nuclear core at $\rho=\rho_{0}$. Therefore, for the $\Sigma\mbox{-}N$ interaction, we consider $U_{\Sigma}^{(N)}=-30$ MeV, as used in Ref. \cite{Malik:2021nas}. Besides, the $\Xi\mbox{-}N$ interaction in nuclear matter is attractive with the potential $U_{\Xi}^{(N)}=-18$ MeV \cite{Char:2014cja}. We take then such a value in our calculation. The $g_{\sigma K}$ can be related to the potential of Kaon at the saturated density through Eq. (\ref{eq3}). $g_{\sigma K}$ values corresponding to several values of $U_{K}$ are listed in Table \ref{tab:Table 1.}.

\begin{table}
\caption{\label{tab:Table 1.}
$g_{\sigma K}$ determined for several $U_{K}$ values in the FSUGold model.
}
\begin{ruledtabular}
\begin{tabular}{lccr}
\textrm{$U_{K}$MeV} &
\textrm{-120} &
\textrm{-140} &
\textrm{-160} \\
$g_{\sigma K}$ & 0.6030 & 1.1775 & 1.7521 
\end{tabular}
\end{ruledtabular}
\end{table}
\par
By solving the Euler-Lagrangian equation of Kaon we obtain equation of motion: $[\emph{D}_{\mu}\emph{D}^{\mu}+m_{\emph{K}}^{* 2}]K=0$. We can then derive the dispersion relation for the Bose-condensation of $\emph{K}^{-}$, which reads
\begin{equation}
\omega_{K}=m_{K}-g_{\sigma K}\sigma-g_{\omega K}\omega-\frac{g_{\rho K}}{2}\rho.
\end{equation}
With the increase of density, the energy $\omega_{K}$ of a test Kaon in the pure normal phase can be computed as a function of the nucleon density. The Kaon energy will decrease while the potential of Kaon($\omega_{K}=\mu_{e}$) increases with the density. When the condition $\omega_{K}=\mu_{e}$ is achieved, the Kaon will occupy a small fraction of the total volume. The new meson field equations are then different from normal phase due to the additional source terms and can be written explicitly as
\begin{equation}
\begin{aligned}
&m_{\sigma}^{2}\sigma+\frac{1}{2}{\kappa}g_{\sigma N}^{3}{\sigma}^{2}+\frac{1}{6}{\lambda}g_{\sigma N}^{4}{\sigma}^{3}=\sum_{B}g_{\sigma B}{\rho}_{B}^{S}+g_{\sigma K}{\rho}_{K}\\
&m_{\omega}^{2}{\omega}+\frac{\xi}{6}g_{\omega N}^{4}{\omega}^{3}+2{\Lambda_{\nu}}g_{\rho N}^{2}g_{\omega N}^{2}{\rho}^{2}{\omega}=\sum_{B}g_{\omega B}{\rho}_{B}-\\
&g_{\omega K}{\rho}_{K}\\
&m_{\rho}^{2}{\rho}+2{\Lambda}_{\nu}g_{\rho N}^{2}g_{\omega N}^{2}{\omega}^{2}{\rho}=\sum_{B}{g_{\rho B}}{\tau}_{3B}{\rho}_{B}-{\frac{g_{\rho K}}{2}{\rho}_{K}},
\end{aligned}
\end{equation}
where $\rho_{N}$ and $\rho_{N}^{S}$ are the nucleon density and the scalar density, respectively and the Kaon density $\rho_{K}=2(\omega_{K}+g_{\omega K}{\omega}+\frac{g_{\rho K}}{2}{\rho})K^{*}K$.

For the neutron matter with baryons and charged leptons, the $\beta$-equilibrium conditions are guaranteed with the following relations of chemical potentials for different particles:
\begin{equation}
\mu_{p}=\mu_{\Sigma^{+}}=\mu_{n}-\mu_{e},
\end{equation}
\begin{equation}
\mu_{\Lambda}=\mu_{\Sigma^{0}}=\mu_{\Xi^{0}}=\mu_{n},
\end{equation}
\begin{equation}
\mu_{\Sigma^{-}}=\mu_{\Xi^{-}}=\mu_{n}+\mu_{e},
\end{equation}
\begin{equation}
\mu_{\mu}=\mu_{e},
\end{equation}
and the charge neutrality condition is fulfilled by:
\begin{equation}
n_{p}+n_{\Sigma}=n_{e}+n_{\mu^{-}}+n_{\Sigma^{-}}+n_{\Xi^{-}}.
\end{equation}
The chemical potential of baryons and leptons read:
\begin{equation}
\mu_{B}=\sqrt{k_{F}^{B 2}+m_{B}^{* 2}}+g_{\omega B}\omega+g_{\rho B}\tau_{3 B} \rho,
\end{equation}
\begin{equation}
\mu_{l}=\sqrt{k_{F}^{l 2}+m_{l}^{2}},
\end{equation}
where $k_{F}^{B}$ is the Fermi momentum and the $m_{B}^{*}$ is the effective mass of baryon B, which can be related to the scalar meson field as $m_{B}^{*}=m_{B}-g_{\sigma B}\sigma$, and $k_{F}^{l}$ is the Fermi momentum of the lepton $l$($\mu$,e). 
\par
The total energy density of the system with Kaon condensation reads then $\varepsilon=\varepsilon_{N}+\varepsilon_{K}$, where $\varepsilon_{N}$ is the energy density of normal nuclear matter and can be given as 
\begin{equation}
\begin{aligned}
\varepsilon_{B}=&\sum_{B}\frac{2}{(2\pi)^{3}}\int_{0}^{K_{F}^{B}}\sqrt{m_{B}^{*}+k^{2}}d^{3}k+\frac{1}{2}m_{\omega}^{2}\omega^{2}\\
&+\frac{\xi}{8}g_{\omega N}^{4}\omega^{4}+\frac{1}{2}m_{\sigma}^{2}\sigma^{2}+\frac{\kappa}{6}g_{\sigma N}^{3}\sigma^{3}+\frac{\lambda}{24}g_{\sigma N}^{4}\sigma^{4}\\
&+\frac{1}{2}m_{\rho}^{2}\rho^{2}+3\Lambda_{\nu}g_{\rho N}^{2}g_{\omega N}^{2}\omega^{2}\rho^{2}+\\
&\frac{1}{\pi^{2}}\sum_{l}\int_{0}^{k_{F}^{l}}\sqrt{k^{2}+m_{l}^{2}}k^{2}dk,
\label{eq6}
\end{aligned}
\end{equation}
where $k_{F}^{B}$ is the Fermi momentum and $m_{B}^{*}$ is the effective mass of baryons, which can be related to the scalar meson field as $m_{B}^{*}=m_{B}-g_{\sigma B}\sigma$. And the energy contributed by the Kaon condensation $\varepsilon_{K}$ is 
\begin{equation}
\varepsilon_{K}=2m_{K}^{* 2}K^{*}K=m_{K}^{*}\rho_{K}.
\end{equation}
The Kaon does not contribute directly to the pressure as it is a (s-wave) Bose condensate so that the expression of pressure reads
\begin{equation}
\begin{aligned}
P=&\sum_{B}\frac{1}{3}\frac{2}{(2\pi)^{3}}\int_{0}^{K_{F}^{B}}\frac{k^{2}}{\sqrt{m_{B}^{\star 2}+k^{2}}}dk^{3}+\frac{1}{2}m_{\omega}^{2}\omega^{2}\\
&+\frac{\xi}{24}g_{\omega N}^{4}\omega^{4}-\frac{1}{2}m_{\sigma}^{2}\sigma^{2}-\frac{\kappa}{6}g_{\sigma N}^{3}\sigma^{3}-\frac{\lambda}{24}g_{\sigma N}^{4}\sigma{4}\\
&+\frac{1}{2}m_{\rho}^{2}\rho^{2}+\Lambda_{\nu}g_{\rho N}^{2}g_{\omega N}^{2}\omega^{2}\rho^{2}+\\
&\frac{1}{3\pi^{2}}\sum_{l}\int_{0}^{k_{F}^{l}}\frac{k^{4}}{\sqrt{k^{2}+m_{l}^{2}}}dk.
\label{eq8}
\end{aligned}
\end{equation}
With the obtained $\varepsilon$ and $P$, the mass-radius relation and other relevant quantities of neutron star can be obtained by solving the Oppenheimer and Volkoff equation \cite{glendenning2012compact,Baldo:1997ag,oppenheimer1939massive}.
\begin{equation}
\begin{aligned}
\frac{dP(r)}{dr}=&-\frac{GM(r)\varepsilon}{r^{2}}(1+\frac{P}{\varepsilon C^{2}})(1+\frac{4\pi r^{3}P}{M(r)C^{2}})\\
&\times(1-\frac{2GM(r)}{rC^{2}})^{-1},
\end{aligned}
\end{equation}
\begin{equation}
dM(r)=4\pi r^{2}\varepsilon(r)dr,
\end{equation}
where G is the gravitational constant and C is the velocity of light, and the EOS for neutron matter is given by Eq. (\ref{eq6}) and Eq. (\ref{eq8}), we can study the physical behavior of neutron stars for the extended model.

\par
The $\sigma$ -cut scheme \cite{Maslov:2015lma}, which is able to stiffen the EOS above saturation density, adds in the original Lagrangian density, the function \cite{Maslov:2015lma,Dutra:2015hxa,Kolomeitsev:2015qia}
\begin{equation}
\Delta U(\sigma)=\alpha \ln(1+exp[\beta(f-f_{s,core})]),
\end{equation}
where $f=g_{\sigma N}/M_{N}$ and $f_{s,core}=f_{0}+c_{\sigma}(1-f_{0})$. $M_{N}$ is the nucleon mass. $f_{0}$ is the value of $f$ at saturation density, equal to 0.31 for the FSUGold model. $c_{\sigma}$ is a positive parameter that we can adjust. The smaller $c_{\sigma}$ is, the stronger the effect of the $\sigma$-cut scheme becomes. However, we must be careful that this scheme would not affect the saturation properties of nuclear matter. In this paper, we want to find a suitable value for the parameter$c_{\sigma}$ that is able to satisfy the maximum mass constraint. $\alpha$ and $\beta$ are constants, taken to be $4.822 \times 10^{-4}M_{N}^{4}$ and 120 as in Ref. \cite{Maslov:2015lma}. This scheme stiffens the EOS by quenching the decreasing of the effective mass of the nucleon $M_{N}^{*}=M_{N}(1-f)$ at high density.
\par
Before giving our numerical results, we list parameters for the FSUGold model in Table \ref{tab:Table 2.}. The parameter of the models can be found in Ref. \cite{Fattoyev:2010tb,Fattoyev:2010rx,Fattoyev:2010mx} in detail.
\begin{table*}
\caption{\label{tab:Table 2.}
Parameter sets for the FSUGold model discussed in the text and the meson masses $M_{\sigma}=491.5 MeV$, $M_{\omega}=782.5 MeV$, $M_{\rho}=763 MeV$.
}
\begin{ruledtabular}
\begin{tabular}{lccccccr}
\textrm{Model}&
\textrm{$g_{\sigma}^{2}$} &
\textrm{$g_{\omega}^{2}$} &
\textrm{$g_{\rho}^{2}$} &
\textrm{$\kappa$} &
\textrm{$\lambda$} &
\textrm{$\xi$} &
\textrm{$\Lambda_{\nu}$}\\
FSUGold & 112.19 & 204.54 & 138.47 & 1.42 & 0.0238 & 0.06 & 0.03/0\\
\end{tabular}
\end{ruledtabular}
\end{table*}
\section{\label{sec:level3}Results}
\begin{figure}
\centering
\includegraphics{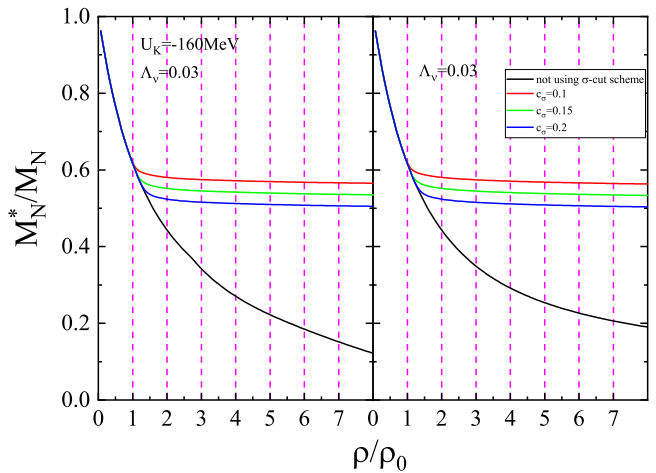}
\caption{Effective mass of nucleons versus baryon density in NS matter using and not using $\sigma$-cut scheme. Left panel: $n$, $p$, leptons, hyperons, $K^{-}$, right panel: only $n$ , $p$, leptons.}
\label{fig1}
\end{figure}
First,we want to find the range for $c_{\sigma}$ in which the saturation properties of nucleon matter are not affected by the $\sigma$-cut scheme, by examining the effective mass of nucleons under the $\sigma$-cut scheme. In Fig. \ref{fig1}, we plot the ratio of the effective mass to the rest mass as a function of the baryon density, where $\rho_{0}$ is the saturation density 0.148 $fm^{-3}$. We can see that, when $\rho \leq \rho_{0}$, the effect mass is almost same as nucleons-only matter and unchanged by the $\sigma$-cut scheme, when $\rho$\textgreater$\rho_{0}$, the effect mass dropped to around $0.55 M_{N}$. 
\begin{figure}
\centering
\includegraphics{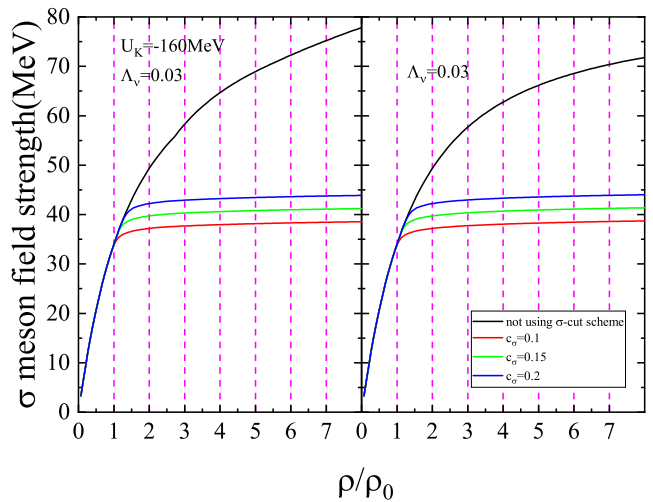}
\caption{$\sigma$ meson field strength as a function of baryon density in NS matter using and not using $\sigma$-cut scheme. Left panel: $n$, $p$, leptons, hyperons, $K^{-}$, right panel: only $n$ , $p$, leptons.}
\label{fig2}
\end{figure}
\par
In Fig. \ref{fig2}, we plot the $\sigma$ meson field strength as a function of the baryon density with and without the $\sigma$-cut scheme. we must make sure that the $\sigma$ potential is not affected by the $\sigma$-cut scheme at saturation density. It is clear that when the baryon density below the saturation density, the $\sigma$ meson field strength is almost same as nucleons-only matter and unchanged by $\sigma$\mbox{-}cut scheme. So we can conclude that, for $\rho$ \textgreater $\rho_{0}$, the $\sigma$ meson field strength is quenched at high baryon density, and the $\sigma$-cut scheme will not affect the saturation properties. This is what we want by using the $\sigma$-cut scheme. The smaller $c_{\sigma}$ is, the stronger the quenching becomes.
\begin{figure}
\centering
\includegraphics{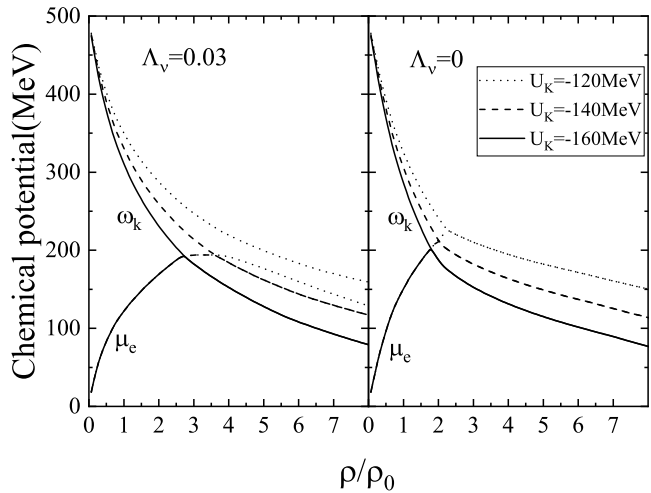}
\caption{Kaon energy($\omega_{k}$) and electron chemical potential($u_{e}$) as a function of baryon density, left panel:$\Lambda_{\nu}=0.03$, right panel:$\Lambda_{\nu}=0$. The solid curve exhibits $U_{K}=-160$ MeV, dotted lines exhibits $U_{K}=-140$ MeV, dashed lines exhibits $U_{K}=-120$ MeV.}
\label{fig3}
\end{figure}
\par
Fig. \ref{fig3} shows the Kaon energy($\omega_{k}$) and electron chemical potential($u_{e}$) as a function of baryon density with $U_{K}=-120,-140,-160$ MeV for the parameter $\Lambda_{\nu}=0$ and $\Lambda_{\nu}=0.03$. $K^{-}$ condensation initiates once the value of $\omega_{K}$ reaches that of the electron chemical potential. For $\Lambda_{\nu}=0.03$, there is no intersection when $U_{K}=-120$ MeV.
\begin{figure}
\centering
\includegraphics{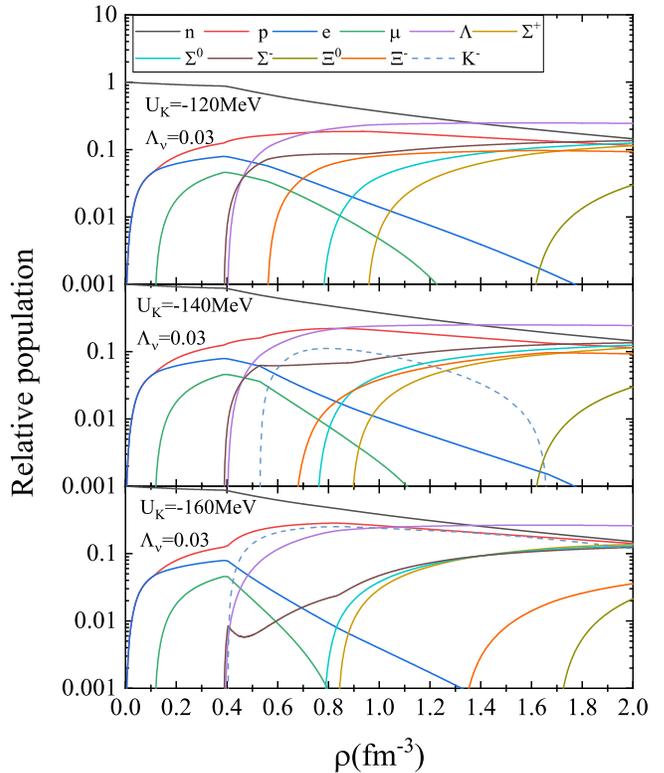}
\caption{Relative population of particles versus baryon density without $\sigma$-cut scheme and $K^{-}$ potential depth of $U_{K}=-120, 140 ,160$ MeV, $\Lambda_{\nu}=0.03$, dashed lines denote $K^{-}$.}
\label{fig4}
\end{figure}
\par
Fig. \ref{fig4} shows the relative population of particles versus baryon density with Kaon optical potential $U_{K}=-120,-140,-160$ MeV, $\Lambda_{\nu}=0.03$. When $U_{K}=-120$ MeV, there is no $K^{-}$. For $U_{K}=-140$ MeV, the mixed phase initiates with the onset of $K^{-}$ at $\sim 3.5 \rho_{0}$ and terminates at $\sim 8.7 \rho_{0}$, for $U_{K}=-160$ MeV, with the appearance of $K^{-}$ at $\sim 2.7 \rho_{0}$ and ceasing of electron population around $\sim 8.9 \rho_{0}$, the proton and $K^{-}$ population becomes equal following the charge neutrality condition. $\Sigma^{-}$ will decrease, also $\Xi^{-}$ will be delayed.
\par
\begin{figure}
\centering
\includegraphics{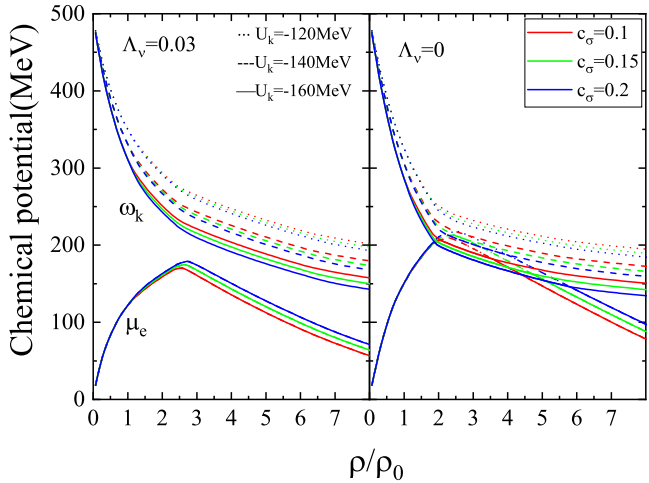}
\caption{Kaon energy($\omega_{k}$) and electron chemical potential($u_{e}$) as functions of baryon density with the $\sigma$-cut scheme, left panel: $\Lambda_{\nu}$=0.03, right panel: $\Lambda_{\nu}$=0. The solid curve exhibits $U_{K}=-160$ MeV, dotted lines exhibits $U_{K}=-140$ MeV, dashed lines exhibits $U_{K}=-120$ MeV.}
\label{fig5}
\end{figure}

 
Next we examine the effect of the $\sigma$-cut scheme on the Kaon energies. This is plotted in the Fig. \ref{fig5}. For $\Lambda_{\nu}=0.03$, there is no intersection between $\omega_{K}$ and $\mu_{e}$. The bigger $c_{\sigma}$ is, the closer $\omega_{K}$ and $\mu_{e}$ are. For $\Lambda_{\nu}=0$, the Kaon condensation will occur when $U_{K}=-140,-160$ MeV. The smaller $c_{\sigma}$ is, the smaller $\sigma$ meson field strength becomes, this will reduce the appearance of $K^{-}$. The isoscalar-isovector coupling ($\Lambda_{\nu}$) term in Eq. (\ref{eq:one}) is used to modify the density dependence of the symmetry energy and the neutron skin thicknesses of heavy nuclei, we can clearly find that $\Lambda_{\nu}$ has a significant impact on the appearance of $K^{-}$. When $\Lambda_{\nu}=0$, this will increase the chemical potential of the electron and promote the appearance of $K^{-}$. 
\begin{figure}
\includegraphics{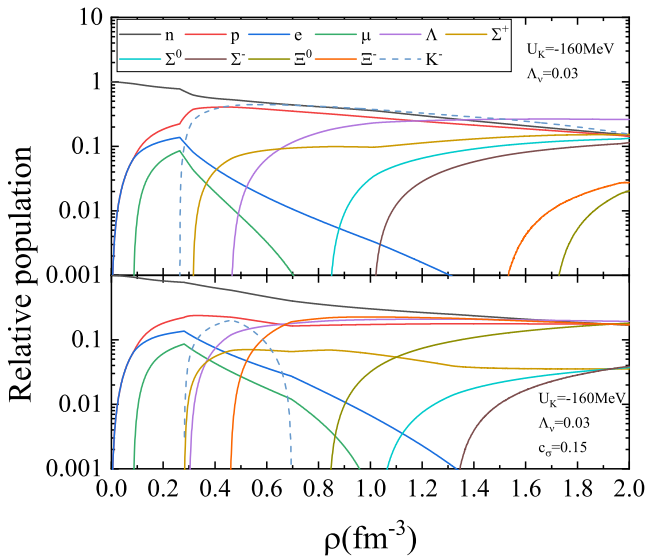}
\caption{Relative population of particles versus baryon density with and without $\sigma$-cut scheme, $K^{-}$ optical potential depth of $U_{K}=-160$ MeV, $\Lambda_{\nu}=0$. Upper panel: without $\sigma$-cut scheme, lower panel: $c_{\sigma}=0.15$, dashed lines denote $K^{-}$.}
\label{fig6}
\end{figure}
\par 
It is interesting to know how the $\sigma$-cut scheme affects the relative populations of particles. In Fig. \ref{fig6}, the relative populations as a function of the baryon density are plotted with $K^{-}$ optical potential ${U_{K}=-160}$ MeV, we are interested in the occurrence of $K^{-}$, so we choose ${\Lambda_{\nu}}=0$. We can find that the percentages of $K^{-}$ is decreased by the $\sigma$-cut scheme, and the relative populations of some hyperons are influenced significantly by the $\sigma$-cut scheme when $\Lambda_{\nu}=0$. The $\sigma$-cut scheme increases the percentage of $\Xi^{-}$ and $\Xi^{0}$, while it decreases the percentage of $\Sigma^{-}$ and $\Sigma^{0}$. We list the threshold densities $n_{cr}$ for Kaon condensation for different values of $K^{-}$ optical potential depths $U_{K}$ in Table \ref{tab:Table 3.}.
\begin{table}
\caption{\label{tab:Table 3.}
Threshold densities $n_{cr}$ (in units of $fm^{-3}$) for Kaon condensation in dense nuclear matter for different values of $K^{-}$ optical potential depths $U_{K}$ (in units of MeV) without $\sigma\mbox{-}$cut scheme and $c_{\sigma}=0.15$.
}
\begin{ruledtabular}
\begin{tabular}{c|cc|cc}
\multirow{2}{*}{$U_{K}$ (MeV)} & \multicolumn{2}{c|}{$n_{cr}(K^{-})$(no $\sigma\mbox{-}$cut scheme)} & \multicolumn{2}{c}{$n_{cr}(K^{-})$($c_{\sigma}=0.15$)} \\
 & \multicolumn{1}{c}{$\Lambda_{\nu}=0.03$} & $\Lambda_{\nu}=0$ & \multicolumn{1}{c}{$\Lambda_{\nu}=0.03$} & $\Lambda_{\nu}=0$ \\ \hline
-120 & \multicolumn{1}{c}{$none$} & 0.34 & \multicolumn{1}{c}{$none$} & $none$ \\ 
-140 & \multicolumn{1}{c}{0.54} & 0.3 & \multicolumn{1}{c}{$none$} & 0.36 \\ 
-160 & \multicolumn{1}{c}{0.41} & 0.27 & \multicolumn{1}{c}{$none$} & 0.28 \\
\end{tabular}
\end{ruledtabular}
\end{table}

\begin{figure}
\centering
\includegraphics{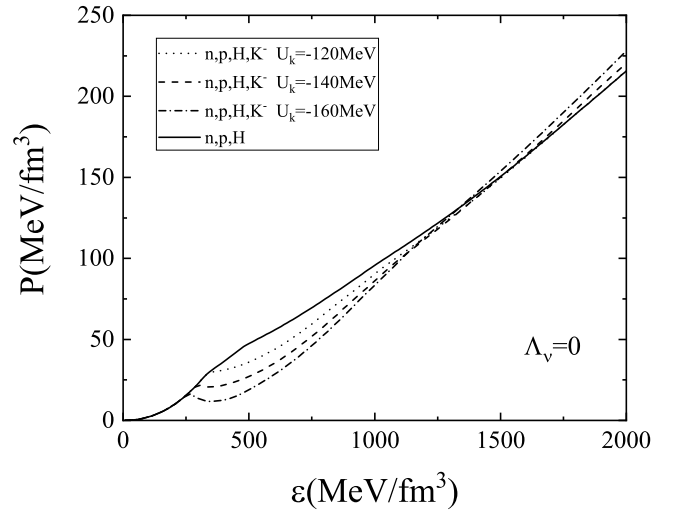}
\caption{Pressure versus energy density without the $\sigma$-cut scheme. The solid line is for n, p, leptons and hyperons whereas others are with additional $K^{-}$, dotted line exhibits $U_{K}=-120$ MeV, dashed line exhibits $U_{K}=-140$ MeV, dash-dotted line exhibits $U_{K}=-160$ MeV.}
\label{fig7}
\end{figure}
\begin{figure}
\centering
\includegraphics{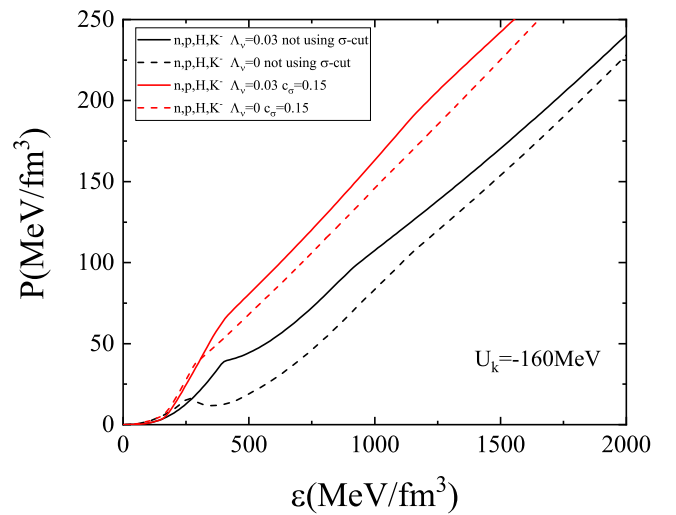}
\caption{Pressure versus energy density with and without the $\sigma$-cut scheme. The solid lines curve with $\Lambda_{\nu}=0.03$, dashed lines curve with $\Lambda_{\nu}=0$, the black lines exhibits without $\sigma$-cut scheme, the red lines exhibits with $c_{\sigma}=0.15$.}  
\label{fig8}
\end{figure}
\par
Then we can discuss some properties of the neutron star, Fig. \ref{fig7} shows the matter pressure as a function of energy density for the Kaon optical potential $U_{K}=-120,-140,-160$ MeV and without $\sigma$-cut scheme. The appearance of Kaon to a great extent softens the EOS. As the energy density increases, the EOS of NS matter that contains K mesons will exceed that of only nucleons and hyperons.
\par
The parameter $\Lambda_{\nu}$ describing the interaction between $\rho$ meson and $\omega$ meson in the FSUGold model is introduced to soften the symmetry energy. It can affect the macroscopic properties of the neutron stars. In Fig. \ref{fig8} the EOS is displayed for $\Lambda_{\nu}=0$ and $\Lambda_{\nu}=0.03$. We can find that the EOS with parameter $\Lambda_{\nu}=0$ is softer than the case of $\Lambda_{\nu}=0.03$, by using $\sigma$-cut scheme, the EOS is significantly stiffened when $c_{\sigma}=0.15$.
\par
\begin{table*}[htb]
\caption{\label{Table4.}
The maximum mass(in unit of solar mass $M_{\odot}$) and radius(in unit of km) of neutron stars using and not using $\sigma$-cut scheme.}
\begin{ruledtabular}
\begin{tabular}{c|cc|cc|cc|cc}
\multirow{2}{*}{} & \multicolumn{2}{c|}{$\Lambda_{\nu}=0.03$} & \multicolumn{2}{c|}{$\Lambda_{\nu}=0$} & \multicolumn{2}{c|}{MSP J0740+6620} & \multicolumn{2}{c}{PSR J0030-0451} \\  
 & \multicolumn{1}{c}{M} & R & \multicolumn{1}{c}{M} & R & \multicolumn{1}{c}{M} & R & \multicolumn{1}{c}{M} & R \\ \hline
no $c_{\sigma}$\mbox{-}cut scheme($n,p,H$) & \multicolumn{1}{c}{1.31} & 11.6 & \multicolumn{1}{c}{1.32} & 13.3 & \multicolumn{1}{c}{\multirow{4}{*}{$2.08\pm0.07$}} & \multirow{4}{*}{$12.39^{+1.3}_{-0.98}$} & \multicolumn{1}{c}{\multirow{4}{*}{$1.34^{+0.15}_{-0.16}$}} & \multirow{4}{*}{$12.71^{+1.14}_{-1.19}$} \\ 
$c_{\sigma}=0.1$($n,p,H,K^{-}$) & \multicolumn{1}{c}{2.07} & 14.1 & \multicolumn{1}{c}{2.07} & 15.1 & \multicolumn{1}{c}{} &  & \multicolumn{1}{c}{} &  \\ 
$c_{\sigma}=0.15$($n,p,H,K^{-}$) & \multicolumn{1}{c}{1.99} & 13.6 & \multicolumn{1}{c}{1.96} & 14.4 & \multicolumn{1}{c}{} &  & \multicolumn{1}{c}{} &  \\ 
$c_{\sigma}=0.2$($n,p,H,K^{-}$) & \multicolumn{1}{c}{1.9} & 13.1 & \multicolumn{1}{c}{1.86} & 13.7 & \multicolumn{1}{c}{} &  & \multicolumn{1}{c}{} &  \\ 
\end{tabular}
\end{ruledtabular}
\end{table*}
\begin{figure}
\centering
\includegraphics{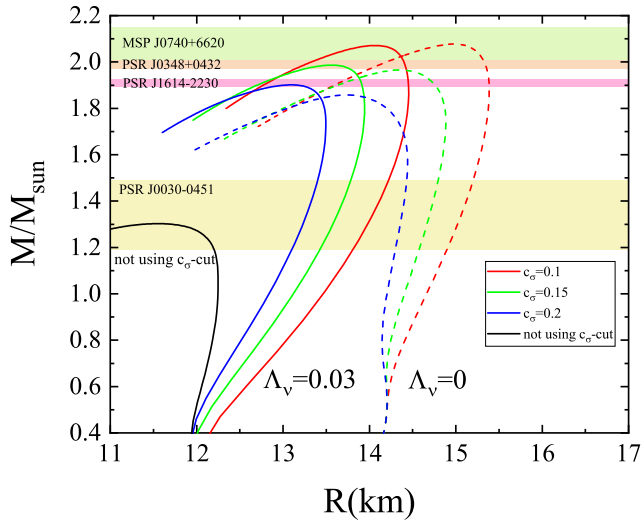}
\caption{Mass-radius relation using and not using $\sigma$-cut scheme in NS matter including hyperons and $K^{-}$. The solid lines denoted $\Lambda_{\nu}=0.03$, dashed lines denoted $\Lambda_{\nu}=0$, $U_{K}=-160$ MeV. The horizontal bars indicate the observational constraints of PSR J1614 - 2230 \cite{Demorest:2010bx,ozel2010massive}, PSR J0348 + 0432 \cite{antoniadis2013j}, MSP J0740 + 6620 \cite{fonseca2021refined,cromartie2020relativistic} and PSR J0030-0451 \cite{Riley_2019}.}
\label{fig9}
\end{figure}
\par
The results of mass-radius relation for static spherical stars from solution of the Tolman-Oppenheimer-Volkoff(TOV) equation discussed here are shown in Fig. \ref{fig9}. The mass measurements of PSR J1614 - 2230 \cite{Demorest:2010bx,arzoumanian2018nanograv,fonseca2016nanograv,ozel2010massive}, PSR J0348 + 0432 \cite{antoniadis2013j}, MSP J0740 + 6620 and PSR J0030 - 0451 are indicated by the horizontal bars. We find that this scheme can significantly increase the maximum mass of the neutron star, the smaller $c_{\sigma}$ is, the stronger the effect of this scheme is. The parameter $\Lambda_{\nu}$ will not significantly affect the maximum mass of the neutron star, but will increases the radius, note that there is no appearance of $K^{-}$ when $\Lambda_{\nu}=0.03$ from Fig. \ref{fig5}. We list the simultaneous measurement of radius for MSP J0740 + 6620 \cite{fonseca2021refined,cromartie2020relativistic} and PSR J0030 - 0451 \cite{Riley_2019} by the NICER data and maximum mass of the neutron star for various values of $c_{\sigma}$ and $\Lambda_{\nu}$ in Table \ref{Table4.}.
\section{\label{sec:level4}Summary}
In this paper, we have discussed the $K^{-}$ meson condensation inside the neutron star under the FSUGold model. This model predicts a limiting neutron star mass of $1.72M_{\odot}$ \cite{ToddRutel:2005zz}. The maximum mass changes to $1.3M_{\odot}$ when hyperons are included \cite{Wu:2011zzb}. Adding Kaon meson will further soften the EOS. In this work we used $\sigma$-cut scheme, by adjusting the parameter $c_{\sigma}$, we got the maximum mass heavier than $2M_{\odot}$ within the range of observed mass measurements. We also compared the radius of NS with the recent NICER results, the $\rho\mbox{-}\omega$ coupling constant $\Lambda_{\nu}$ has a significant effect on the radius the appearance of $K^{-}$ and when using $\sigma$-cut scheme. The Kaon condensation cannot exist in the NS matter for $\Lambda_{\nu}=0.03$, for $\Lambda_{\nu}=0$, as the $\rho-\omega$ interaction is switched off, the electron chemical potential will increase and make it possible for $K^{-}$ condensation occur in neutron star.
\bibliography{prc}
\bibliographystyle{unsrt}
\end{document}